\documentclass[aps,prd,superscriptaddress,showpacs,preprint]{revtex4}
\usepackage{graphicx, bm}

\begin{document}
\draft
\title{Limits on the quartic couplings $Z\gamma\gamma\gamma$ and $ZZ\gamma\gamma$ from $e^+e^-$ colliders}

\author{A. Guti\'errez-Rodr\'{\i}guez}
\affiliation{\small Facultad de F\'{\i}sica, Universidad Aut\'onoma de Zacatecas\\
             Apartado Postal C-580, 98060 Zacatecas, M\'exico.\\}

\author{C. G. Honorato}
\affiliation{\small Departamento de F\'{\i}sica, CINVESTAV.\\
             Apartado Postal 14-740, 07000, M\'exico D.F., M\'exico.\\}

\author{J. Monta\~no}
\affiliation{\small Departamento de F\'{\i}sica, CINVESTAV.\\
             Apartado Postal 14-740, 07000, M\'exico D.F., M\'exico.\\}

\author{M. A. P\'erez}
\affiliation{\small Departamento de F\'{\i}sica, CINVESTAV.\\
             Apartado Postal 14-740, 07000, M\'exico D.F., M\'exico.\\}

\date{\today}

\begin{abstract}

We obtain limits on the quartic neutral gauge bosons couplings
$Z\gamma\gamma\gamma$ and $ZZ\gamma\gamma$ using LEP 2 data published
by the L3 Collaboration on the reactions $e^+e^-\to \gamma\gamma\gamma, Z\gamma\gamma$.
We also obtain $95\hspace{0.8mm}\%$ C. L. limits on these couplings at the future linear colliders
energies. The LEP 2 data induce limits of order $10^{-5}\hspace{0.8mm}GeV^{-4}$ for the $Z\gamma\gamma\gamma$
couplings and of order $10^{-3}\hspace{0.8mm}GeV^{-2}$ for the $ZZ\gamma\gamma$ couplings, which are still
above the respective Standard Model predictions. Future $e^+e^-$ linear colliders may improve
these limits by one or two orders of magnitude.
\end{abstract}

\pacs{14.70.-e, 13.85.Lg, 13.85.Rm
\vspace*{2cm}
\noindent  E-mail: $^{1}$alexgu@fisica.uaz.edu.mx, $^{2}$mperez@fis.cinvestav.mx}

\vspace{5mm}

\maketitle


\section{Introduction}

Neutral gauge bosons self couplings provide a window to study physics
beyond the Standard Model (SM) \cite{Ellison,Barroso,Hernandez}. While trilinear neutral gauge
boson couplings (TNGC) $V_iV_jV_k$, with $V_i=Z,\gamma$, test the gauge
structure of the SM \cite{Hernandez}, it has been argued that quartic neutral gauge
boson couplings (QNGC) $V_iV_jV_kV_l$ may provide a connection to the mechanism
of electroweak symmetry breaking \cite{Ellison}. Since the longitudinal
components of the $Z$ gauge boson are Goldstone bosons associated to the
electroweak symmetry breaking mechanism, these QNGC could represent then
a connection with the scalar sector of the gauge theory that has become popular
after the recent evidence of a new boson with a mass around $125\hspace{0.8mm}GeV$ \cite{ATLAS}.
However, it has been found recently in a detailed calculation of the one-loop induced decay mode
$Z\to \gamma\gamma\gamma$, in both the SM and the 331 model, that the respective
scalar contributions are suppressed with respect to the dominant
virtual fermionic contributions \cite{Montaño}. This is also the case in the
one-loop contributions to TNGC \cite{Hernandez,Belanger}. The QNGC are induced by effective
operators of dimension greater or equal to six and, in the SM, the QNGC are highly
suppresed, with the only exception of the $ZZZZ$ vertex, because they arise at the
one-loop level \cite{Belanger,M.A.Perez}. Any deviation from the SM predictions for the
QNGC will be associated to a signal of new physics effects \cite{Ellison}.

While considerable effort has been devoted to study the TNGC,
the QNGC are only starting to receive some attention. TNGC have
been measured with an accuracy of the few percent level at
LEP 2 \cite{Villa} and the Tevatron \cite{Abazov}, while QNGC are only loosely
constrained at LEP 2 \cite{Villa}.
In fact, the $Z\gamma\gamma\gamma$ couplings have not been
bounded yet by direct measurements \cite{Villa}. In the present paper, we are
interested in obtaining limits on the quartic vertices
$Z\gamma\gamma\gamma$ and $ZZ\gamma\gamma$ coming from the LEP 2 data on the reactions
$e^+e^-\to \gamma\gamma\gamma, Z\gamma\gamma$ that were used to get limits on the anomalous
$HZ\gamma$ coupling but not on the QNGC \cite{Achard,Gutierrez}. We will obtain also $95 \%$ C. L.
limits on these quartic couplings at the future International Linear Collider (ILC) and the
Compact Linear Collider (CLIC) \cite{ILC,Brau:2012hv}. Since there is not a published account,
as far as we know, of the calculation of the $Z\gamma\gamma\gamma$ vertices in the SM, we present
a brief analysis on the connection of the $Z\gamma\gamma\gamma$ form factors to the analytical results
obtained in Ref. \cite{Montaño} for the branching ratio of the decay mode $Z\gamma\gamma\gamma$ in both
the SM and the 331 model. However, a similar calculation for the $ZZ\gamma\gamma$ form factors in the SM
is not available in the published literature.

Constraints on the anomalous quartic gauge couplings $ZZ\gamma\gamma$ have been studied in
$\gamma\gamma$ and $Z\gamma$ fusion processes at the LHC \cite{Chapon}, in $ZZ\gamma$, $Z\gamma\gamma$
production processes at future $e^+e^-$ linear colliders \cite{Belanger} and from
the non observation of the rare decay $Z\to \nu\bar\nu\gamma\gamma$ at LEP 1 \cite{M.A.Perez}.
However, constraints on the anomalous $Z\gamma\gamma\gamma$ vertex are more difficult to get.
In the present paper we find that the negative search for the reactions
$e^+e^- \to \gamma\gamma\gamma, Z\gamma\gamma$ at LEP 2 by the L3 Collaboration may be
translated into limits of order $10^{-5}\hspace{0.8mm}GeV^{-4}$ on the $Z\gamma\gamma\gamma$ couplings and of
order $10^{-3}\hspace{0.8mm}GeV^{-2}$ on the $ZZ\gamma\gamma$ couplings. We also find that sensitivity studies
on these couplings at future $e^+e^-$ colliders may improve these limits by one or two orders
of magnitude.

The paper is organized as follows. In Section II we present the calculation of the respective
cross sections for the processes $e^+e^- \to \gamma\gamma\gamma, Z\gamma\gamma$ and in Section III
we include our results and conclusions. In particular, we present the connection among our quartic
couplings $G_{1, 2}$ and the results obtained in Ref. \cite{Montaño} for the branching ratio of
the decay mode $Z\to \gamma\gamma\gamma$ in the SM and the 331 model.

\section{Cross Sections}

We will use the following parameterizations for the QNGC \cite{Barroso,Stohr},

\begin{eqnarray}\label{Effective-Z3gamma}
{\cal L}_{Z\gamma\gamma\gamma}&=&\frac{G_1}{\Lambda^4}F_{\mu\nu}F^{\mu\nu}F_{\rho\sigma}Z^{\rho\sigma}
+\frac{G_2}{\Lambda^4}F_{\mu\nu}F^{\mu\rho}F_{\rho\sigma}Z^{\sigma\nu},\label{z3g-effec}\\
{\cal L}_{ZZ\gamma\gamma}&=&-\frac{e^2}{16\Lambda^2c^2_W}a_0F_{\mu\nu}F^{\mu\nu}Z^\alpha Z_\alpha
-\frac{e^2}{16\Lambda^2c^2_W}a_c F_{\mu\nu}F^{\mu\alpha}Z^\nu Z_\alpha,\label{2z2g-effec}
\end{eqnarray}

\noindent where $F_{\mu\nu}=\partial_\mu A_\nu-\partial_\nu A_\mu$ and $Z_{\mu\nu}=\partial_\mu Z_\nu-\partial_\nu Z_\mu$
are the respective gauge tensor fields for the photon and the $Z$ boson. $\Lambda$ represents the energy scale at which
new physics interactions may appear. The respective Feynman rules for these effective vertices are thus given by,

\begin{eqnarray}\label{z3g-vert}
&&\sum^6_{a=1}iP_a\left\{\frac{G_1}{\Lambda^4}\left[\frac{}{}(p_1\cdot p_2)
(p_2\cdot p_3)g_{\alpha\rho}g_{\mu\nu}- (p_1\cdot p_3)p_{1\nu} p_{2\mu}
g_{\alpha\rho}-(p_1\cdot p_3)p_{1\nu} p_{2\alpha} g_{\rho \mu}+p_{1\nu} p_{1\rho}
p_{2\mu} p_{3\alpha} \right]\right. \nonumber\\
&&+\left. \frac{G_2}{\Lambda^4}\left[-\frac{}{}(p_1\cdot p_2) (p_1\cdot
p_3)g_{\alpha\mu}g_{\rho\nu} + (p_2\cdot p_3)p_{1\alpha}p_{1\nu}
g_{\rho\mu}-(p_2\cdot p_3)p_{1\nu} p_{1\rho}
g_{\alpha \mu} + (p_2\cdot p_3)p_{1\nu} p_{2\alpha} g_{\rho\mu}\right. \right.\nonumber\\
&&\left.\left. +2(p_2\cdot p_3)p_{1\rho} p_{2\mu} g_{\alpha \nu} - (p_1\cdot
p_3)p_{1\alpha} p_{2\rho} g_{\mu\nu} - p_{1\alpha} p_{1\nu} p_{2\rho}
p_{3\mu}\frac{}{} \right]\right\},
\end{eqnarray}

\noindent and

\begin{eqnarray}\label{2z2g-vert}
&&\frac{ie^2}{8c^2_W\Lambda^2}\left\{4a_0g^{\alpha\beta}\left[\frac{}{}(p_1\cdot
p_2)g^{\mu\nu}-p_1^\nu p_2^\mu\right] +a_c
\left[\frac{}{}(p_1^\alpha p_2^\beta + p_1^\beta
p_2^\alpha)g^{\mu\nu} + (p_1\cdot p_2)(g^{\mu\alpha} g^{\nu\beta}+
g^{\nu\alpha}g^{\mu\beta})\right.\right.\nonumber\\
&&-\left.\left. p_1^\nu(p_2^\beta g^{\mu\alpha}+p_2^\alpha
g^{\mu\beta}) - p_2^\mu(p_1^\beta g^{\nu\alpha}+p_1^\alpha
g^{\nu\beta})\frac{}{}\right]
\right\},
\end{eqnarray}

\noindent where the four momenta $p_{1,2,3}$ correspond to the emitted
photons and $P_a$ denotes possible permutations
$(p_1, \mu) \longleftrightarrow (p_2, \nu)\longleftrightarrow (p_3, \rho)$.

All the couplings $G_{1, 2}$ and $a_{0, c}$ are CP conserving and within
the SM all of them vanish at tree level.
As far as we know, the $a_{0, c}$ have not been computed explicitly in the SM,
whereas the couplings $G_{1, 2}$ can be extracted directly from the recent
calculation performed in the SM and the 331 model \cite{Pisano} for the branching
ratio of the rare decay mode $Z\to \gamma\gamma\gamma$ \cite{Montaño}. Since these
authors did not use explicitly the parametrization given in Eqs. (\ref{z3g-effec})
and (\ref{2z2g-effec}), we include the connection of the $G_{1, 2}$ couplings to
the results obtained for the $Z\to \gamma\gamma\gamma$ decay in Ref. \cite{Montaño}.
These form factors are dominated by the fermionic virtual contributions and they are
essentially the same in both the SM and the 331 model, but unfortunately with rather
low values, $1.63\times 10^{-10}$ and $1.33\times 10^{-10}$, respectively.

\begin{figure}[!ht]
\centerline{\scalebox{0.74}{\includegraphics{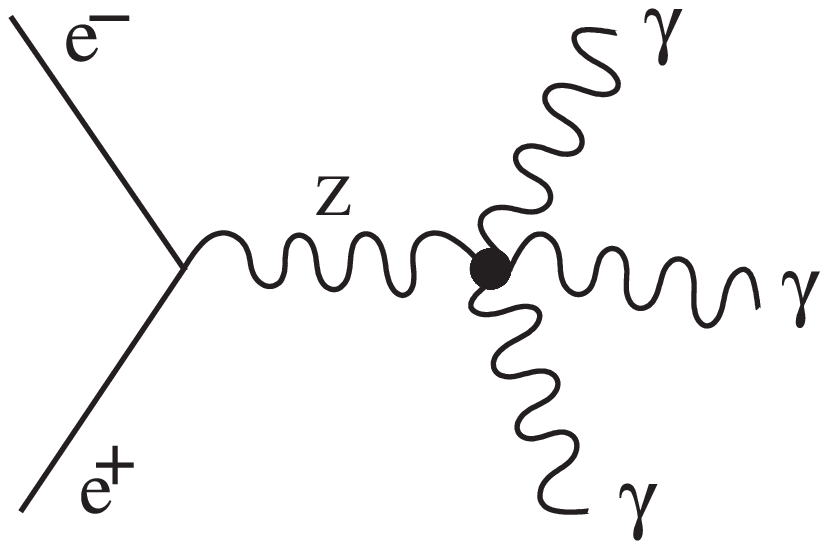}}}
\caption{ \label{fig:z3gamma} Feynman diagram for the process
$e^+e^-\to \gamma\gamma\gamma$ induced by the effective vertex
$Z\gamma\gamma\gamma$.}
\end{figure}

\begin{figure}[!ht]
\centerline{\scalebox{0.74}{\includegraphics{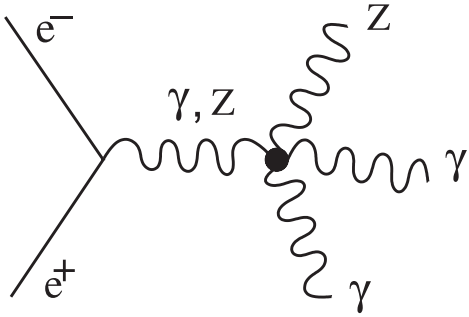}}}
\caption{ \label{fig:2z2gamma} Feynman diagrams for the process
$e^+e^-\to Z\gamma\gamma$ induced by the effective vertices
$ZZ\gamma\gamma$ and $Z\gamma\gamma\gamma$.}
\end{figure}

In Figures \ref{fig:z3gamma} and \ref{fig:2z2gamma} we present the contributions of the
effective interactions given in Eqs. (\ref{z3g-vert}) and (\ref{2z2g-vert}) to the processes
$e^{+}e^{-}\rightarrow \gamma\gamma\gamma$ and $e^{+}e^{-}\rightarrow Z\gamma\gamma$. The SM
contributions to these processes occur via t-channel diagrams involving initial-state radiation
\cite{Barroso,Belanger}. The SM cross section for the process $e^+e^-\to Z\gamma\gamma$ has been
computed by Stirling and Werthenbach \cite{Belanger} for energies greater than $200\hspace{1mm}GeV$.
In Figure 3 we present the SM results for the cross sections of both processes, they are of
order of few femtobarns as it was obtained in Ref. \cite{Belanger} for the $Z\gamma\gamma$ case.
According to this reference, in order to reduce the contributions due to initial-state radiation
in these reactions, the L3 Collaboration introduced cuts on the photon energies and their polar
angles, $E_\gamma > 5$ $GeV$ and $|\cos\theta_\gamma|< 0.97$ \cite{Achard}. Events from
$e^{+}e^{-}\rightarrow \gamma\gamma\gamma, Z\gamma\gamma$ processes were selected by requiring
the photon candidates to lay in the central region of the detector with $|\cos\theta_\gamma|<0.8$
and a total CM electromagnetic energy large than $\sqrt{s}/2$. In this case, the L3 Collaboration
was interested in getting limits on the anomalous Higgs couplings $HZ\gamma$ and $H\gamma\gamma$.
However, using their data we are able to get also limits on the $G_{1, 2}$ and $a_{0, c}$
couplings: $G_1/\Lambda^4 < 1.2\times 10^{-5}\hspace{0.8mm}GeV^{-4}$,
$G_2/\Lambda^4 < 9.4\times 10^{-6}\hspace{0.8mm}GeV^{-4}$,
$a_0/\Lambda^2 < 5.9\times 10^{-3}\hspace{0.8mm}GeV^{-2}$ and
$a_c /\Lambda^2< 1.6\times 10^{-2}\hspace{0.8mm}GeV^{-2}$. The latter limits are close to the
bounds obtained by the L3 Collaboration from a direct search of $Z\gamma\gamma$ events at LEP 2
energies: [-0.008, 0.021]\hspace{0.8mm}$GeV^{-2}$ and [-0.029, 0.039]\hspace{0.8mm}$GeV^{-2}$,
respectively \cite{Villa}.

The expressions for the respective cross sections induced by
the effective vertices given in Eqs. (\ref{z3g-vert}) and (\ref{2z2g-vert}) are given by

\begin{equation}\label{cross-Z3g}
\sigma(e^+e^-\to \gamma\gamma\gamma)= \frac{\alpha
M^{10}_Z}{1105920 \pi^2} \left[
\frac{1-4x_W+8x^2_W}{x_W(1-x_W)}\right]
\left[\frac{2\left(\frac{G_1}{\Lambda^4}\right)^2+3\left(\frac{G_2}{\Lambda^4}\right)^2-3\left(\frac{G_1}{\Lambda^4} \right)\left( \frac{G_2}{\Lambda^4}\right)}{(s-M^2_Z)^2+M^2_Z\Gamma^2_Z}\right],
\end{equation}

\noindent and

\begin{eqnarray}\label{cross-2Z2g}
\sigma(e^+e^-\to Z\gamma\gamma)&=&\frac{\alpha M^6_Z(s-M^2_Z)^4}{5308416 \pi^2 s^4}
\left[71\left(\frac{G_1}{\Lambda^4}\right)^2 + 138\left(\frac{G_1}{\Lambda^4}\right)\left(\frac{G_2}{\Lambda^4}\right) + 96\left(\frac{G_2}{\Lambda^4}\right)^2\right]\nonumber\\
&+&\frac{5\alpha M^6_Z(s-M^2_Z)^4}{4608 \pi^2 s^4 \left[(s-M^2_Z)^2+M^2_Z\Gamma^2_Z\right]}
\left[ \frac{1-4x_W+8x^2_W}{x_W(1-x_W)}\right]\nonumber\\
&\times&\left[\frac{a_0^2}{\Lambda^4}+\frac{1}{8}\frac{a_c^2}{\Lambda^4}
+\frac{1}{2}\frac{a_c}{\Lambda^2}\frac{a_0}{\Lambda^2}\right].
\end{eqnarray}

In Eq. (\ref{cross-2Z2g}), the first term comes from the Feynman diagrams shown in
Fig. \ref{fig:2z2gamma} for the exchanged photon and the second one comes from the
exchanged $Z$ boson. We did not include the contribution coming from the interference
of the two diagrams because we will get limits on the form factors one at the time.

\section{RESULTS AND CONCLUSIONS}

In order to get limits on the $G_{1,2}$ quartic couplings, we shall use the bound
obtained by the L3 Collaboration on the cross section $\sigma(e^+e^- \to Z \to \gamma\gamma\gamma)$
at LEP 2 energies \cite{Achard} and our expression for this cross section in terms of
the quartic couplings given in Eq. (5). Similarly, we have used the bounds on the cross
section $\sigma(e^+e^-\to Z\gamma\gamma)$ obtained by the L3 Collaboration in order to
get limits on the $a_{0,c}$ couplings specified in Eq. (6). Since this expression does not
improve the limits on the quartic couplings $G_{1,2}$, in this case we have set them to
zero in order to obtain the limits on the $a_{0,c}$ couplings. We obtain in this approach
$G_1/\Lambda^4 < 1.2\times 10^{-5}\hspace{0.8mm}GeV^{-4}$, $G_2/\Lambda^4 < 9.4\times 10^{-6}\hspace{0.8mm}GeV^{-4}$,
$a_0/\Lambda^2 < 5.9\times 10^{-3}\hspace{0.8mm}GeV^{-2}$ and $a_c /\Lambda^2< 1.6\times 10^{-2}\hspace{0.8mm}GeV^{-2}$.
The respective $95\hspace{0.8mm}\%$ sensitivity limits for the $a_{0,c}$ couplings will be
obtained for CM energies of 500 and 1000\hspace{0.8mm}$GeV$ and the luminosity
expected at the ILC/CLIC accelerators: 500\hspace{0.8mm}$fb^{-1}$.

\begin{figure}[!ht]
\centerline{\scalebox{0.53}{\includegraphics{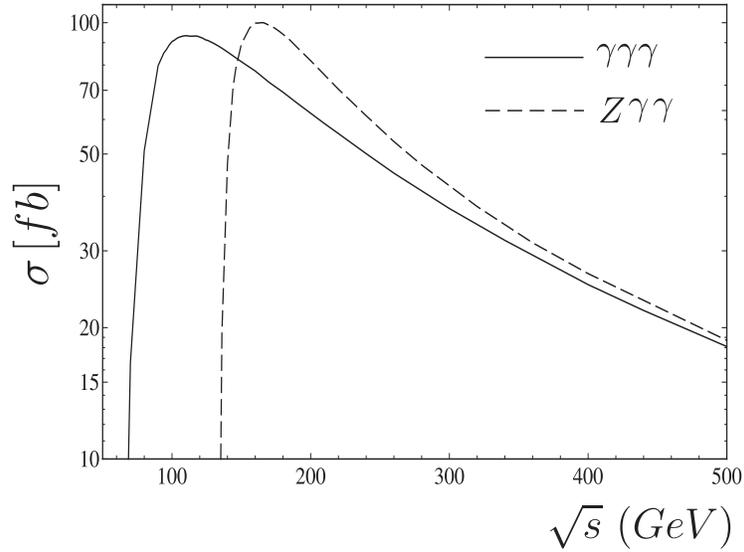}}}
\caption{ \label{fig:Z3gammas-2Z2gamma} Cross sections for the processes
$e^+e^-\to \gamma\gamma\gamma, Z\gamma\gamma$ as function of the CM
energy in the SM. We have used cuts on the photon energies and their
polar angles, $E_\gamma > 20\hspace{0.8mm}GeV$ and $|\cos\theta_\gamma| < 0.8$.}
\end{figure}

In both cases we shall rely on the respective SM predictions for their cross sections.
In Figure \ref{fig:Z3gammas-2Z2gamma} we depict the dependence of these cross sections with
respect to the CM energy. We have used cuts on the photon energies and their polar angles:
$E_\gamma > 20\hspace{0.8mm}GeV$ and $|\cos\theta_\gamma| < 0.8$. In the case of the production
of the three photons, our result agrees with that obtained by Stirling and Werthenbach \cite{Belanger}.
On the other hand, the Feynman diagrams shown in Figures 1 and 2 for the effective vertices
contributions generate an increase for the cross sections with respect to the CM energy that
may dominate the SM contributions. We will use this tendency in order to get $95\% \hspace{0.8mm}C. L.$
limits on these effective vertices for future $e^+e^-$ colliders.

\begin{figure}[!ht]
\centerline{\scalebox{0.515}{\includegraphics{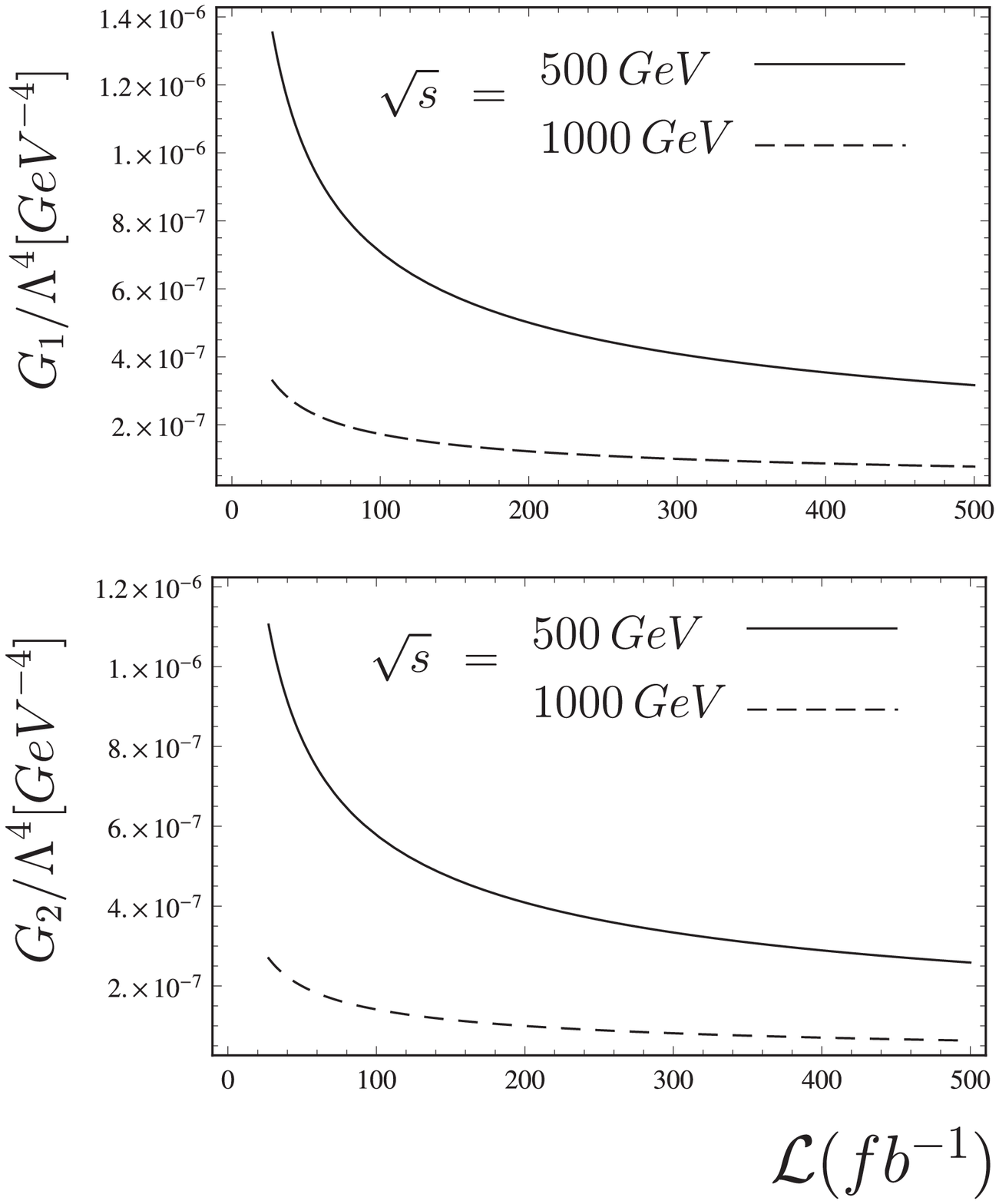}}}
\caption{ \label{fig:sen-z3gamma} Sensitivity limits at $95\hspace{0.8mm}\%$ C.L.
for the couplings $G_{1, 2}/\Lambda^4\hspace{0.8mm}[GeV^{-4}]$ as function of the integrated luminosity
for two ILC/CLIC CM energies. We have taken the $G_{1, 2}$ couplings one at the time.}
\end{figure}

\begin{figure}[!ht]
\centerline{\scalebox{0.41}{\includegraphics{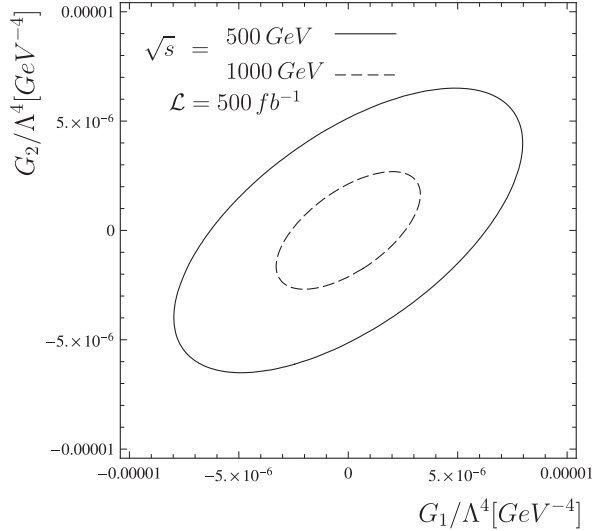}}}
\caption{ \label{fig:coun-Z3gamma} Contours limits at $95\hspace{0.8mm}\%$ C. L. in
the $G_1$-$G_2$ plane for the process $e^+e^-\to \gamma\gamma\gamma$
for $\sqrt{s}=500, 1000$\hspace{0.8mm}$GeV$ and ${\cal L}=500$\hspace{0.5mm}$fb^{-1}$.
We have taken the $G_{1,2 }$ couplings simultaneously.}
\end{figure}

\begin{figure}[!ht]
\centerline{\scalebox{0.519}{\includegraphics{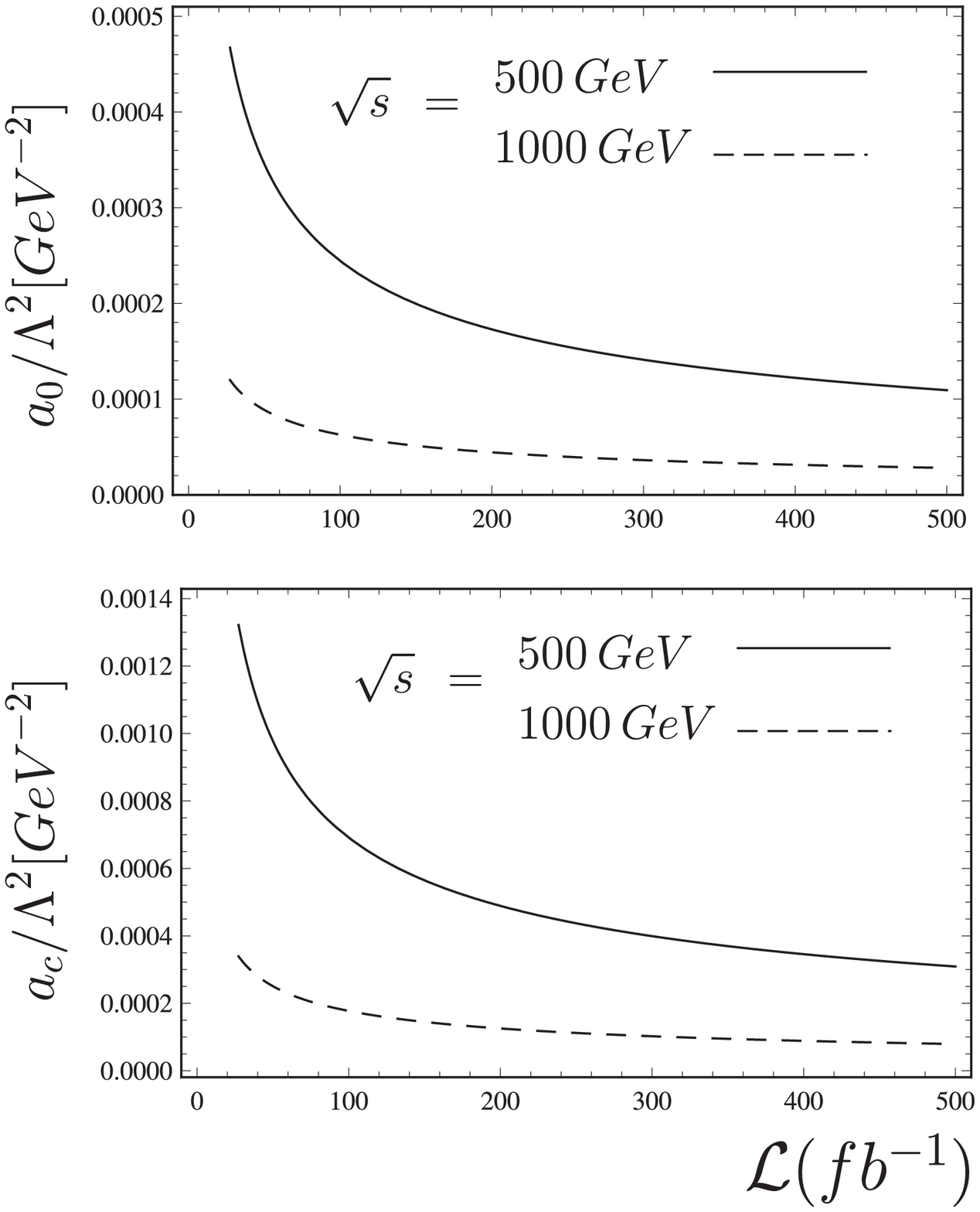}}}
\caption{ \label{fig:sen-2Z2gamma} Sensitivity limits at $95\hspace{0.8mm}\%$ C.L.
for the couplings $a_{0, c}/\Lambda^2\hspace{0.8mm}[GeV^{-2}]$ as function of the integrated luminosity
for two ILC/CLIC CM energies. We have taken the $a_{0, c}$ couplings one at the time.}
\end{figure}

\begin{figure}[!ht]
\centerline{\scalebox{0.41}{\includegraphics{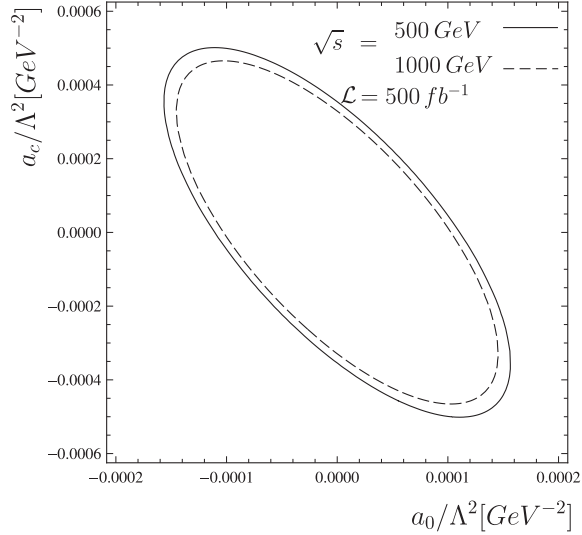}}}
\caption{ \label{fig:con-2Z2gamma} Contours limits at $95\hspace{0.8mm}\%$ C. L. in
the $a_0$-$a_c$ plane for the process $e^+e^-\to Z\gamma\gamma$
for $\sqrt{s}=500, 1000$\hspace{0.2mm}$GeV$ and ${\cal L}=500$\hspace{0.5mm}$fb^{-1}$.
We have taken the $a_{0,c }$ couplings simultaneously.}
\end{figure}

Using the numerical values $\sin^2\theta_W=x_W=0.2314$,
$M_{Z}=91.18$\hspace{0.8mm}$GeV$ and $\Gamma_{Z}=2.49$\hspace{0.8mm}$GeV$ \cite{PDG2012}, we
obtain the cross sections for the processes $e^+e^-\to \gamma\gamma \gamma, Z\gamma\gamma$ as
functions of the CM energy and the $G_{1, 2}, a_{0, c}$ couplings. We have also implemented
in our calculation the cut used by the L3 Collaboration on the photons energy and their
polar angle in order to suppress the SM contributions associated to initial-state radiation.
In Fig. \ref{fig:sen-z3gamma} we depict the sensitivity limits at $95 \%\hspace{0.8mm} C. L.$
for the $G_{1, 2}$ couplings for CM energies of 500\hspace{0.8mm}$GeV$ and 1000\hspace{0.8mm}$GeV$
and we have taken the $G_{1, 2}$ couplings one at the time. The respective combined limits contours
are shown in Fig. \ref{fig:coun-Z3gamma}. On the other hand, in order to get sensitivity limits and
the respective limits contours for the $a_{0, c}$ couplings we have set to zero the contribution
associated to the $G_{1, 2}$ couplings in Eq. (\ref{cross-2Z2g}). The respective limits are
given in Figures \ref{fig:sen-2Z2gamma} and \ref{fig:con-2Z2gamma} also for two different
values of the energy of the ILC and CLIC accelerators.

In Figures \ref{fig:sen-z3gamma} and \ref{fig:sen-2Z2gamma} we have used the statistical significance
expression given in terms of the expected number of signal and background events in the reactions
$e^+e^-\to \gamma\gamma\gamma, Z\gamma\gamma$ \cite{Han}. We have assumed that the background events
arise from the SM contributions while the signal events come from the effective vertices contributions
shown in Figures 1 and 2. In order to obtain sensible limits on the effective vertices, we also assume
that the SM contribution is smaller than the new physics contributions. The number of expected events
then are given by the integrated luminosity and the respective cross section. In figures 5 and 7 we used
two CM energies planned for the ILC/CLIC accelerators in order to get $95\% \hspace{1mm}C. L.$ contour
limits for the $Z\gamma\gamma\gamma$ and $ZZ\gamma\gamma$ effective couplings and the planned luminosity
of $500\hspace{1mm}fb^{-1}$. We can appreciate that these limits are about two orders of magnitude lower
with respect to those obtained from the data obtained by the L3 Collaboration.

In order to compare our results for the quartic couplings with the SM predictions, we use the results
obtained in Ref. \cite{Montaño} for the branching ratio of the decay $Z\to \gamma\gamma\gamma$. A complete
one-loop calculation of the Feynman diagrams for this decay mode was presented for the SM and the 331 model \cite{Pisano}.
The relation of our quartic couplings $G_{1,2}$ to the $F_{Z_i}$ form factors used in this reference is given by

\begin{equation}
\left(\frac{G_{1,2}}{\Lambda^4}\right)^2 = 2\bigg[\frac{8\alpha^2(M_Z)}{s_Wc_W}\bigg]^2
    \int_0^1\int_{1-x}^1\Big|F_{ZG_{1,2}}^{\frac{1}{2}}+F_{ZG_{1,2}}^1+F_{ZG_{1,2}}^0\Big|^2 dydx \ ,
\end{equation}

\noindent where $x$ and $y$ are kinematical variables associated to the $Z\to \gamma\gamma\gamma$
decay mode \cite{Montaño} and $g = e/s_W = \sqrt{4\pi\alpha}/s_W$. Each $F_{Z_i}$ form
factor identifies the fermionic, vectorial and scalar contributions to the one-loop diagrams.
It was found that the dominant contribution comes from the fermionic amplitudes. In Table I
we include the predictions expected for the $G_{1,2}$ quartic couplings for the SM, 331 model and
from the PDG limits for the respective branching ratio using the expression for the decay width

\begin{equation}\label{Width}
\Gamma(Z\to \gamma\gamma\gamma)=\frac{M_Z^9}{552960\pi^3} \frac{(2G_1^2+3G_2^2-3G_1G_2)}{\Lambda^8}.
\end{equation}

\begin{table}[!ht]
\caption{\label{Gi-Br-effective-table} Values of $G_{1,2}/\Lambda^4$ [$GeV^{-4}$] as function of BR according to
Eq. (16) of Refs. \cite{Stohr,V0-3g-331}. We include the PDG 2012 limit for BR$(Z\to \gamma\gamma\gamma)$ \cite{PDG2012}.}
\begin{center}
 \begin{tabular}{lcc}
\hline\hline
BR                          & $|G_1/\Lambda^4 \hspace{0.8mm}[GeV^{-4}]|$  & $|G_2/\Lambda^4 \hspace{0.8mm}[GeV^{-4}]|$ \\
\hline
$10^{-5}$(PDG)              & 2.22$\times10^{-8}$  & 1.81$\times10^{-8}$ \\
5.41$\times10^{-10}$(SM)    & 1.63$\times10^{-10}$ & 1.33$\times10^{-10}$ \\
5.26$\times10^{-10}$(331) & 1.61$\times10^{-10}$ & 1.31$\times10^{-10}$ \\
\hline\hline
\end{tabular}
\end{center}
\end{table}

In conclusion, we have obtained limits on the quartic couplings $Z\gamma\gamma\gamma$ and
$ZZ\gamma\gamma$ at LEP 2 energies by using published L3 data for the reactions
$e^+e^-\to \gamma\gamma \gamma, Z\gamma\gamma$. Our limits obtained from the LEP 2
data on the reaction $e^+e^-\to Z\gamma\gamma$ are close to the best limits obtained in the
LEP collider \cite{Villa}. In this case, SM predictions for the $a_{0,c}$ couplings are not
available in the literature. Our $95\%$ sensitivity limits expected for these couplings at
ILC/CLIC energies are of order $10^{-4}\hspace{1mm}GeV^{-2}$ for a luminosty of $500\hspace{1mm}fb^{-1}$.
These limits are close to those obtained by Stirling and Werthenbach for a $300\hspace{1mm}fb^{-1}$
luminosity \cite{Belanger}. Similar limits have been obtained from the process $e^+e^-\to \nu\bar\nu \gamma\gamma$
\cite{Montagna} and through effects induced by the polarization of the $Z$ gauge boson and initial
state radiation in the process $e^+e^-\to Z\gamma\gamma$ \cite{Baillargeon}.

\vspace{0.8cm}

\begin{center}
{\bf Acknowledgments}
\end{center}

We acknowledge support from CONACyT, SNI and PROMEP (M\'exico).

\newpage

\newpage

\newpage

%

\newpage

\begin{thebibliography}{99}

\bibitem{Ellison} J. Ellison and J. Wudka, {\it Annu. Rev. Nucl. Part. Sci.} {\bf 4}8, 33 (1998);
                  G. Weiglein {\it et al.}, LHC/ILC Study Group, {\it Phys. Rept.} {\bf 426}, 47 (2006);
                  S. Godfrey, AIP Conf. Proc. 350, 41 (1995); arXiv:hep-ph/9505252;
                  J. J. Toscano, {\it AIP Conf. Proc.} {\bf 857B}, 103 (2006).

\bibitem{Barroso} A. Barroso {\it et al.}, {\it Z. Phys.} {\bf C28}, 149 (1985).

\bibitem{Hernandez} J. M. Hern\'andez, {\it et al.}, {\it Phys. Rev.} {\bf D60}, 013004 (1999);
                    G. J. Gounaris, {\it et al.}, {\it Phys. Rev.} {\bf D62}, 073013 (2000);
                    F. Larios, {\it et al.}, {\it Phys. Rev.} {\bf D63}, 113014 (2001);
                    M. A. P\'erez, G. Tavares Velasco and J. J. Toscano, {\it Int. J. Mod. Phys.} {\bf A19}, 159 (2004);
                    O. Cata, arXiv:1304.1008 [hep-ph].

\bibitem{ATLAS} G. Aad, {\it et al.}, ATLAS Collaboration, {\it Phys. Lett.} {\bf B716}, 1 (2012);
                S. Chatrchyan, {\it et al.}, CMS Collaboration, {\it ibid.} {\bf 30}, (2012).

\bibitem{Montaño} J. Monta\~no, {\it et al.}, {\it Phys. Rev.} {\bf D85}, 035012 (2012);
                  A. Denner, {\it et al.}, {\it Eur. Phys. J.} {\bf C20}, 201 (2001).

\bibitem{Belanger} G. Belanger, F. Boudjema, {\it Phys. Lett.} {\bf B288}, 201 (1992);
                   W. J. Stirling, A. Werthenbach, {\it Eur. Phys. J.} {\bf C14}, 103 (2000);
                   G. Montagna, {\it et al.}, {\it Nucl. Phys.} {\bf B541}, 31 (1999).

\bibitem{M.A.Perez} M. A. P\'erez, G. Tavares-Velasco, and J. J. Toscano, {\it Phys. Rev.} {\bf D67}, 017702 (2003).

\bibitem{Villa} S. Villa, {\it Nucl. Phys.} {\bf B} (Proc. Suppl.) {\bf 142}, 391 (2005), and references therein.

\bibitem{Abazov} V. M. Abasov, {\it et al.}, D0 Collaboration, {\it Phys. Lett.} {\bf B653}, 378 (2007);
                 D. Acosta, {\it et al.}, CDF Collaboration, {\it Phys. Rev. Lett.} {\bf 94}, 041803 (2005).

\bibitem{Achard} P. Achard {\it et al.}, L3 Collaboration, {\it Phys. Lett.} {\bf B589}, 89 (2004);
                 {\it ibid}, {\it Phys. Lett.} {\bf B540}, 43 (2002).

\bibitem{Gutierrez} A. Guti\'errez-Rodr\'iguez, J. Monta\~no and M. A. P\'erez, {\it J. Phys. G: Nucl. Part. Phys.} {\bf G38}, 095003 (2011).

\bibitem{ILC} T. Abe, {\it et al.}, American Linear Collider Group, hep-ex/0106057;
              J. A. Aguilar-Saavedra, {\it et al.}, ECFA/DESY Lc Physics Working Group, hep-ph/0106315;
              Koh Abe, {\it et al.}, ACFA Linear Collider Working Group, hep-ph/0109166;
              ILC Technical Review Committee, second report, 2003, SLAC-R-606, February 2003;
              E. Accomando, {\it et al.}, CLIC Physics Working Group, hep-ph/0412251.

\bibitem{Brau:2012hv} J. E. Abreu, {\it et al.}, arXiv:1210.0202 [hep-ex].

\bibitem{Chapon} E. Chapon, C. Royon, O. Kepka, {\it Phys. Rev.} {\bf D81}, 074003 (2010);
                 I. Sahin and B. Sahin, {\it Phys. Rev.} {\bf D86}, 115001 (2012);
                 R. S. Gupta, {\it Phys. Rev.} {\bf D85}, 014006 (2012).

\bibitem{Stohr} M. Stohr and J. Horejs\'i, {\it Phys. Rev.} {\bf D49}, 3775 (1994);
                J. Horejs\'i, M. Stohr, {\it Z. Phys.} {\bf C64}, 407 (1994).

\bibitem{Pisano} F. Pisano and V. Pleitez, {\it Phys. Rev.} {\bf D46}, 410 (1992);
                 H. Frampton, {\it Phys. Rev. Lett.} {\bf 69}, 2889 (1992).

\bibitem{PDG2012} J. Beringer, {\it et al.}, Particle Data Group, {\it Phys. Rev.} {\bf D86}, 010001 (2012).

\bibitem{Han} T. Han and J. Jiang, {\it Phys. Lett.} {\bf B516}, 337 (2001).

\bibitem{V0-3g-331} A. Flores-Tlalpa, J. Monta\~no, F. Ramirez-Zavaleta and J. J. Toscano, {\it Phys. Rev.} {\bf D80}, 033006 (2009).

\bibitem{Montagna} G. Montagna, {\it et al.}, {\it Phys. Lett.} {\bf B515}, 197 (2001).

\bibitem{Baillargeon} M. Baillargeon {\it et al.}, {\it Z. Phys.} {\bf C71}, 431 (1996).

\end{thebibliography}
\end{document}